
\magnification=\magstep1
\hsize 6.0 true in
\vsize 9.0 true in
\voffset=-.5truein
\pretolerance=10000
\baselineskip=13truept

\font\tentworm=cmr10 scaled \magstep2
\font\tentwobf=cmbx10 scaled \magstep2

\font\tenonerm=cmr10 scaled \magstep1
\font\tenonebf=cmbx10 scaled \magstep1

\font\eightrm=cmr8
\font\eightit=cmti8
\font\eightbf=cmbx8
\font\eightsl=cmsl8
\font\sevensy=cmsy7
\font\sevenm=cmmi7

\font\twelverm=cmr12
\font\twelvebf=cmbx12
\def\subsection #1\par{\noindent {\bf #1} \noindent \rm}

\def\mid {\let\rm=\tenonerm \let\bf=\tenonebf \rm \bf}

\def\para{\par \vskip 12 pt}

\def\head{\let\rm=\tentworm \let\bf=\tentwobf \rm \bf}

\def\heading #1 #2\par{\centerline {\head #1} \smallskip
 \centerline {\head #2} \vskip .15 pt \rm}

\def\eight{\let\rm=\eightrm \let\it=\eightit \let\bf=\eightbf
\let\sl=\eightsl \let\sy=\sevensy \let\m=\sevenm \rm}

\def\foots{\noindent \eight \baselineskip=10 true pt \noindent \rm}
\def\sexion{\let\rm=\twelverm \let\bf=\twelvebf \rm \bf}

\def\section #1 #2\par{\vskip 20 pt \noindent {\mid #1} \enspace {\mid #2}
  \para \noindent \rm}

\def\abstract#1\par{\para \foots {\bf Abstract: \enspace}#1 \para}

\def\author#1\par{\centerline {#1} \vskip 0.1 true in \rm}

\def\abstract#1\par{\noindent {\bf Abstract: }#1 \vskip 0.5 true in \rm}

\def\sqr#1#2{{\vcenter{\vbox{\hrule height.#2pt
  \hbox {\vrule width.#2pt height#1pt \kern#1pt
  \vrule width.#2pt}
  \hrule height.#2pt}}}}

\def\n{\noindent}
\def\s{\smallskip}
\def\m{\medskip}
\def\b{\bigskip}
\def\c{\centerline}

\def\gne #1 #2{\ \vphantom{S}^{\raise-0.5pt\hbox{$\scriptstyle #1$}}_
{\raise0.5pt \hbox{$\scriptstyle #2$}}}

\def\ooo #1 #2{\vphantom{S}^{\raise-0.5pt\hbox{$\scriptstyle #1$}}_
{\raise0.5pt \hbox{$\scriptstyle #2$}}}


\voffset=-.5truein
\vsize=9truein
\baselineskip=22pt
\hsize=6.0truein
\pageno=1
\pretolerance=10000
\def\n{\noindent}
\def\s{\smallskip}
\def\b{\bigskip}
\def\m{\medskip}
\def\c{\centerline}
\baselineskip=20pt

\line{\hfill IUCAA - 14/94}

\line{\hfill June 1994}
\m
\m

\c{\bf\mid Uniqueness of the non-singular family and}
\c{\bf\mid characterisation of cosmological models  }

\m
\m
\m
\m
\m
\m
\m
\m
\m
\c{Naresh Dadhich$^1 $}
\c{Inter University  Centre for Astronomy \& Astrophysics}
\c{ P.O. Box 4, Pune-411007, India}
\c {and }
\c{Department of Mathematics \& Applied Mathematics }
\c{ University of Natal, Durban 4001, South Africa.}
\b
\s
\c{L K Patel}
\c{Department of Mathematics, Gujarat University}
\c{Ahmedabad -- 380 009, India}
\m
\m
\m
\m
\m
\m
\m

\c{\bf Abstract}
\s
\indent We prove that
for an orthogonal  spacetime metric separable in space and time in comoving
coordinates, the requirements of perfect fluid and non-singularity single out
the unique family
of singularity free cosmological  models. Further homogeneous models could only
be Bianchi I or FLRW while inhomogeneous ones can be with or without a
singularity.
\m
\m
\m
\m
\m
\m
\m
\m
\m
\m
\n PACS Numbers: 04.20Jb, 98.80Dr.
\s
\s
\s
\s

$^1 $ E-mail address: naresh@iucaa.ernet.in
\vfill\eject
\baselineskip=24pt
\indent The characteristics of the relativistic cosmology are the expansion and
the big-bang singular origin of the Universe. The imprint of the latter is
believed to be seen in the cosmic microwave background radiation observations
[1,2]. The theoretical model imbibing these features is an exact solution of
Einstein's equations known as the Friedman-Lemaitre-Robertson-Walker (FLRW)
model. It represents a homogeneous and isotropic Universe filled with perfect
fluid. This is the generally accepted current model of the Universe. Despite
its success, there are some questions of principles that deserved to be
addressed to. The model is highly specialised for it is homogeneous and
isotropic. These properties could by no means be considered generic enough
features of the Universe. The Universe should in fact have more general initial
conditions. Secondly, inhomogeneity may be quite appropriate for evolution of
large scale structures in the Universe. Above all even at the present e!
 poch, the question of homogeneity
\s
\indent It is therefore important to find solutions of Einstein's equations
when homogeneity and isotropy could be dropped. The first step in this
direction came with the Bianchi models that are homogeneous but anisotropic.
For a long time only homogeneous models were studied. Inhomogeneity came in
through what is known as orthogonally transitive $G_2 $ cosmologies [3-5].
These are the simplest inhomogeneous models that admit two spacelike Killing
vectors which are mutually as well as hypersurface orthogonal. The class of
solutions of Einstein's equations of inhomogeneous family was first obtained by
Wainwright and Goode [6] and subsequently by others [7-9]. Amongst them was a
remarkable solution
obtained by Senovilla [8] which was free of the big-bang (or of any other kind)
singularity. It satisfied all the physical reasonableness conditions and had an
acceptable equation of state $\rho = 3 p > 0 $. Until then on the strength of
the singularity theorems [10] it was generally believed that the occurrence of
big-bang is an inevitable feature of general relativity (GR). The conflict was
soon resolved [11] when it was shown that the solution in question does not
obey the assumption of existence of compact trapped surfaces and hence the
singularity theorems become inapplicable. All prior attempts to construct
singularity free models had either to ascribe physically unacceptable behaviour
for matter leading to violation of energy and causality conditions or to invoke
quantum effects or modification of GR [12,13]. Senovilla's [8] was the first
singularity free solution, true to GR, conforming to energy and causality
conditions.
\s
\indent The most interesting feature of inhomogeneous cosmology is the
non-occurrence of singularity in the models. Hence it has no beginning and no
end. This is aesthetically very attractive and appealing feature, which was
first propounded by the steady state theory. Ruiz and Senovilla [14] have
identified a large family of singularity free cosmological models. This family
is unique for cylindrically symmetric metric separable in space and time in the
comoving coordinates (i.e. fluid lines are orthogonal to $t = const. $
surface). In this note we wish to establish the uniqueness of the non-singular
family by dropping the assumption of cylindrical symmetry and in the process we
succeed in characterising the perfect fluid models.
\s
\indent Before we go any further let us note a general result arising out of
the
following two relations [15],
$$ \theta, \alpha = {3 \over 2}
\bigg[( \sigma_\alpha^i + \omega_\alpha^i)_{;i} - (\sigma_{\alpha i} +
w_{\alpha_i} ) \dot u^i \bigg] \eqno (1) $$

$$ ~~~~= \theta \dot u_{\alpha} + { 1 \over {\sqrt g_{00}}}
\bigg( \ln \sqrt{|g/g_{00}|}\bigg)_{, 0\alpha} \eqno (2) $$

\n where $\theta, \sigma, \omega, \dot{u}_\alpha$ are the
kinematic parameters; expansion, shear, rotation and acceleration,
$\dot{u}_{\alpha} = u_{\alpha; i}u^i$.  We have however assumed
$u_i = \sqrt{g_{00}} \delta_i^0$.
\s
\indent We infer from the above relations:
\s
\noindent {\bf Lemma :} {\it In the absence of shear and vorticity, the
expansion of
fluid is constant over the 3-space orthogonal to the fluid congruence and
further, the acceleration also vanishes when the quantity $g/g_{00}$ is a
separable function of space and time in the comoving coordinates.}
\s
\noindent {\bf Corollary :} {\it For the vorticity free spacetime with the
separability (as is the case for the metric (3)), acceleration can be non-zero
only if shear is non-zero.}
\s
\indent According to the Raychaudhuri equation [16], in the absence of
vorticity acceleration is necessary for
halting
the collapse to avoid singularity which in our case can only exist if shear is
non-zero.  Thus non-singular solutions represented by the metric (3) will
always have to be both inhomogeneous and anisotropic.
\s
\indent In cosmology in general and $G_2 $ cosmologies in particular the metric
is generally taken to be orthogonal and the comoving coordinates are employed.
That is the fluid velocity vector is vorticity free and is orthogonal to $t =
const. $ hypersurface. Further as is common for most cosmological models, we
take the metric to be separable in space and time coordinates. Hence we write
the metric in the form
$$ds^2 = D dt^2 - A dx^2_1 - B dx^2_2 - C dx^2_3 \eqno (3) $$

\n where by sepaarability we mean $A = A(t) A (x_{\alpha}) $ and so on. The
velocity field of the fluid is $U_i = \sqrt D \delta^0_i $.
\s
\indent It can be easily seen that the invariant characterisation of this
metric is given by (i) $\theta, \alpha = \theta \dot u_{\alpha} $ and (ii)
$(\sigma/\theta), \alpha = 0, $ i.e. the anisotropy parameter is constant on
the 3-hypersurface orthogonal to the fluid flow [17]. The condition (i) follows
from eqn (2) while (ii) can be verified easily by writing $\sigma/\theta $ for
the metric (3).
\s
\indent Let us couch the general result as a theorem as follows :
\s
\n {\bf Theorem :} Let the metric (3) be a perfect fluid cosmological solution
of Einstein's equations then it can only be
\s
\item{(a)} if homogeneous, $G_3 $ (homogeneity) - Bianchi I or $G_6 $ (both
homogeneity and isotropy) - FLRW model
\s
\item{(b)} if inhomogeneous, $G_2 $ (admitting only two spacelike Killing
vectors) models with or without the big-bang singularity.
\s
\n Further the singularity free family  as already identified in [14] is unique
and is cylindrically symmetric.
\s
\n {\bf Proof :} The theorem characterises all the perfect fluid models that
the metric (3) can represent. But for the separability and orthogonality of the
metric we make no assumptions. The fluid
conditions alone will impose symmetries on the metric.
\s
\indent The perfect fluid distribution will imply the conditions; $T_{0
\alpha} = 0 ,  T_{\alpha \beta} = 0$ for $\alpha \neq \beta $ and $T_1^1 =
T_2^2 = T_3^3$.
\s
\indent To go any further we need the explicit expressions for $T_i^k$ [18]
which look quite frightening and formidable.  Fortunately, we have discovered
an underlying order in them that allows us to write the rest of them from the
given two (one each of diagonal and off diagonal) by
prescribing the appropriate permutation rules.
We begin by

$$\eqalign{ -32\pi AT_0^1 &= - 2 \bigg({B_0 \over B} +
{C_0 \over C} \bigg)_1 + {A_0 \over A} \bigg({B_1 \over B} +
{C_1 \over C} \bigg) + {B_0 \over B} \bigg(-{B_1 \over B} + {D_1 \over D}\bigg)
\cr
&+ {C_0 \over C} \bigg(- {C_1 \over C} + {D_1 \over D} \bigg) \cr} \eqno (4) $$

$$\eqalign{ -32 \pi T^1_1 &= {1 \over A} \bigg[{B_1 C_1 \over BC} + {D_1 \over
D} \bigg({B_1 \over B} + {C_1 \over C} \bigg) \bigg] \cr
&+ {1 \over B} \bigg[ 2 \bigg({C_2 \over C} + {D_2 \over D} \bigg)_2 + {C_2
\over C} \bigg( - {B_2 \over B} + {C_2 \over C} \bigg) + {D_2 \over D}
\bigg(-{B_2 \over B} + {C_2 \over C} + {D_2 \over D} \bigg) \bigg] \cr
&+ {1 \over C} \bigg[ 2 \bigg({B_3 \over B} + {D_3 \over D} \bigg)_3 + {B_3
\over B} \bigg({B_3 \over B} - {C_3 \over C} \bigg) \cr
&+ {D_3 \over D} \bigg({B_3 \over B} - {C_3 \over C} \bigg) + {D_3 \over D}
\bigg({B_3 \over B} - {C_3 \over C} + {D_3 \over D} \bigg) \bigg] \cr
&+ {1 \over D} \bigg[-2\bigg( {B_0 \over B} + {C_0 \over C} \bigg)_0 - {B_0
\over B} \bigg({B_0 \over B} + {C_0 \over C} - {D_0 \over D} \bigg) - {C_0
\over C} \bigg({C_0 \over C} - {D_0 \over D} \bigg) \bigg] \cr} \eqno (5) $$

\n where a subscript denotes partial
differentiation and here the assumption of separability is not effected.
\s
\indent The successive cyclic permutations $A \rightarrow B \rightarrow C
\rightarrow A $ and $1 \rightarrow 2 \rightarrow 3 \rightarrow 1 $ will give~
$T^2_0, T^3_0 $ ~from ~$T^1_0~; ~ T^2_3, T^3_1 $~ from~ $T^1_2 $;  and~
$T^2_2, T^3_3 $~ from $T^1_1 $. To write $T^1_2 $ from $T^1_0 $, let
$0 \rightarrow i2 $ (i.e. $A_0 \rightarrow iA_2, T^1_0 \rightarrow i T^1_2 $)
and $B \rightarrow C \rightarrow D \rightarrow B $ while $T^0_0 $
follows from $T^1_1 $ for  $2 \rightarrow 3 \rightarrow 1 \rightarrow i0
\rightarrow -2  (T^1_1 \rightarrow T^0_0 $) and $A \rightarrow D  \rightarrow B
\rightarrow C \rightarrow A $. Thus we can write all ten $T_i^k $, given the
two, one each of diagonal and off diagonal.
\s
\indent Let us begin with the general case where the metric is a function of
all the coordinates. Eqns. $T_{\alpha 0} = 0 $, of which $T_{10} = 0 $ reads as

$${D_1 \over D} = \bigg({C_0/C - A_0/A \over B_0/B + C_0/C} \bigg) {C_1 \over
C} + \bigg({B_0/B - A_0/A \over B_0/B + C_0/C} \bigg) {B_1 \over B} \eqno (6)
$$

\n and similarly $D_2/D $ and $D_3/D $. It is clear if $A_0/A = B_0/B = C_0/C $
which implies $\sigma = 0 $ and $\dot u_{\alpha} = 0 $ (because $D_{\alpha }= 0
$ ), then the spacetime is both isotropic and homogeneous and it can be no
other than the big-bang singular FLRW [19].
\s
\indent If $A_0/A \not= B_0/B \not= C_0/C $ and since the metric is assumed to
be separable in $t $ and
$x_{\alpha} $, the above equation will imply

$$C_0/C - A_0/A = k_1 (B_0/B - A_0/A) + n_1 (B_0/B + C_0/C) \eqno (7) $$

and

$$ B_1/B = k_1 C_1/C  \eqno (8)  $$

\n where $k_1 $ and $n_1 $ are constants, and similar equations will follow
from $D_2/D $ and $D_3/D $. We can now set the exact differential for the space
dependence of $D $,
$$ d(ln D) = (ln D)_1 dx_1 + (ln D)_2 dx_2 + (ln D)_3 dx_3 $$

\n which can be integrated along any path to give the same result. Evaluating
it along two different paths, we obtain $B = C^{k_1}, D = C^{n_1} $ and $A =
C^{k_2} $. Thus the space dependence of the metric is all determined but for
the single function $C(x_{\alpha}). $ Eqns. $T_{\alpha \beta} = 0 $ for $\alpha
\not= \beta $ determine $C(x \alpha) = constant $. Hence the metric can only
represent homogeneous and big-bang singular Bianchi I model. When $A_0/A =
B_0/B \not= C_0/C $, eqn (6) and its permutants will imply that either it is
Bianchi I or it admits a spacelike Killing vector. This is the case we consider
next.
\s
\indent We shall now consider the case of metric (3) admitting a spacelike
Killing vector, say ${\partial \over \partial x_3} $. That means the metric
depends only on the two space variables $x_1 $ and $x_2 $.
\s
\indent From (5) it is clear that $T^1_1 = T^2_2 = T^3_3 $ will imply the two
equations of the type,

$${f_1 \over A(t) } + {f_2 \over B(t) } + {f_3 \over C(t)} = F(t) \eqno (9) $$

\n where $f_1, f_2, f_3 $ are functions of space variables $x_{\alpha} $ and
containing derivatives with respect to $x_1, x_2 $ and $x_3 $ respectively.
\s
\indent In this case $ f_3 = 0 $ in (9). Note that $T_{30} \equiv 0 $ now and
$T_{\alpha 0} = 0 $ will lead to $A(t) = B(t) \not= C(t) $ and $D(x_{\alpha}) =
C^k(x_{\alpha}) $, $k $= const. Hence $\sigma $ can be non-zero to give rise to
acceleration which in turn can lead to a viable non-singular case. There will
be three equations ($T_{12} = 0 $ plus the two following from (9)) determining
the space dependence of the metric.
\s
Since $A(t) = B(t) $, it is possible to perform a coordinate transformation to
set $A(x_{\alpha}) = B(x_{\alpha}) $, which means $A = B $. A very detailed and
involved Lie analysis of the equations [20] leads to  the
conclusion that space dependence can only arise as a  function of $x_1 + x_2$
or $x_1^2 + x_2^2$ or $x_1/x_2 $ ( i.e. $A(x_\alpha) = A(x_1+x_2)$ etc).
In either of the first two cases, it could be reduced to the case of
one-coordinate dependence by coordinate transformations. The last case is
obviously singular and would not lead to a viable model.
\s
\n When $A_0/A \not= B_0/B \not= C_0/C $, following the above route, eqns.
$T_{\alpha 0} = 0 $ determine all others in terms of $C(x_{\alpha}), \alpha =
1, 2; T_{12} = 0 $ fixes $C(x_{\alpha}) = (f(x_1) + g(x_2) )^k $ and then eqns.
(9) force $C(x_{\alpha}) $ to be a constant. We are again led to Bianchi I
model.
\s
\indent Finally we come to the case of $G_2 $ models which admit two spacelike
Killing vectors that are mutually as well as hypersurface orthogonal. The
metric depends upon only one space variable.
\s
\indent Now the spacetime is general enough to sustain inhomogeneous perfect
fluid. There will occur two kinds of inhomogeneous fluid models, one with
singularity and the other without it [14]. The singularity free family
possesses cylindrical symmetry and is unique. That means that the already
identified singularity free family [14] is unique not only for cylindrical
symmetry but for the general metric (3). In this case homogeneous models can
however occur but will be Bianchi I only.
\s
\indent Thus is proven the theorem.
\s
\indent We give below the general enough [21] though not the most general [14]
inhomogeneous non-singular cosmological models described by

$$\eqalign{ds^2 & =  \cosh^{2\alpha}(kt) \cosh^{2a} (mr) (dt^2 -
dr^2) - \cosh^{2\beta}(kt) \cosh^{2b}(mr)dz^2  \cr
&-m^{-2}\sinh^2 (mr)\cosh^{2\alpha}(kt)\cosh^{2c}(mr) d\phi^2 \cr} \eqno (10)
$$

\n with $\alpha + \beta = 1$.  It admits the only two cases, (i) $b=c$ and
(ii)$
b+c=1$ for perfect fluid distribution. The former includes the Senovilla's
solution [8] when $c=b=-1/3, a=1, \alpha = 2, \beta =-1$ and $k = 3m$; and
$$ \rho = 3p = {15k^2 \over 8\pi} \cosh ^{-4}(kt) \cosh^{-4}
(kr/3) > 0. \eqno (11) $$
 The latter always gives the stiff fluid,

$$ 8\pi \rho = 8\pi p = (b^2-4)m^2 \cosh^{-2\alpha-2}(2mt)
\cosh^{-2a}(mr) \eqno (12) $$

\n with  $\alpha = 1-b/2, a = b(b-1), c=1-b,
k=2m$.  Clearly $b^2 \geq 4$ which means $b$ must lie outside the interval $-2
\leq b \leq 2$.
\s
Though $\rho > 0$ for both $b>2$ and $b <-2$, the geodesic completeness [22]
demands $b<0$, and hence the latter case is the truly singularity free
solution.  At the two ends of the range for $b = \pm 2$, we have the two
distinct empty space solutions as matter free limits of the stiff fluid.  Here
again, $b=-2$ will be geodesically complete and it can be thought of as
representing field of plane gravitational waves. We have also shown [23]
elsewhere that a natural inhomogenisation of FLRW open model leads to the
non-singular metric (9).
\s
\indent The metric (3) can only describe fluid models of the following kinds;
homogenous FLRW and Bianchi I and $G_2 $  inhomogenous with or without
singularity. That is inhomogeneity cannot be sustained by any symmetry higher
than $G_2 $ [20]. This is an important conclusion; at whatever scale we wish to
incorporate inhomogeneity in fluid models represented by an exact solution of
Einstein's equations, the metric has to be $G_2 $. Howsoever much it is not
supported by observations, it appears essential for bringing in inhomogeneity,
which cannot be ignored even at the present epoch in view of lumpiness in the
Universe. The ideal solution to the question is if inhomogeneity and anisotropy
were scale and time dependent. Then at larger scales and at late times the
model could have evolved to homogeneous and isotropic FLRW. This unfortunately
cannot happen for the metric (3) for $\sigma/\theta $ turns out to be constant
for inhomogeneous fluid models.
\s
\indent It may be noted that the metric (3) is general enough for cosmology.
For both the assumptions of orthogonality and separability are quite common and
shared by most of the cosmological models. It is in this context very important
that it can only admit inhomogeneous perfect fluid models  that could never
isotropise. The only hope is to give up the separability which unfortunately
will make the problem mathematically formidable. If one can succeed in finding
an inhomogeneous non-singular solution of Einstein's equations that isotropises
to approximate to FLRW as time progresses, it will have very important bearing
on our overall cosmological perception of the Universe.
\s
\n{\bf Acknowledgement :} It is a pleasure to thank A.K. Raychaudhuri
for his constructive criticism and discussion. ND thanks Sanjeev Dhurandhar,
Kesh Govinder and
Peter Leach for helpful discussions and the Hanno Rund Fund for a visiting
fellowship.  LKP thanks IUCAA for hospitality that facilitated this work. We
also thank the two referees for their constructive criticism that has resulted
in making the arguments more rigorous.
\vfill\eject
\n{\bf References }
\s
\item{[1]} A.A. Penzias and R.W. Wilson, Ap.J. {\bf 142}, 419 (1965).
\s
\item{[2]} C.F. Smoot et al. Ap.J. Lett. {\bf 396}, L1 (1992).
\s
\item{[3]} C.G. Hewitt and J. Wainwright, Class. Quantum Grav. 7, 2295 (1990).
\s
\item{[4]} J. Wainwright, J. Phys. A {\bf 14}, 1131 (1981).
\s
\item{[5]} M. Carmeli, Ch. Charach, and S. Malin, Phys. Rep. {\bf 76}, 80
(1981).
\s
\item{[6]} J. Wainwright and S.W. Goode, Phys. Rev. {\bf D 22}, 1906 (1980).
\s
\item{[7]} A. Feinstein and J.M.M. Senovilla, Class. Quantum Grav. {\bf 6}, L89
(1989).
\s
\item{[8]} J.M.M. Senovilla, Phys. Rev. Lett. {\bf 64}, 2219 (1990).
\s
\item{[9]} W. Davidson, J. Math. Phys. {\bf 32}, 1560 (1991).
\s
\item{[10]} S.W. Hawking and G.F.R. Ellis, The Large Scale Structure of the
Universe, Cambridge University Press, 1973.
\s
\item{[11]} F.J. Chinea, L. Fernandez-Jambrina, and J.M.M. Senovilla, Phys.
Rev. {\bf D 45}, 481 (1992).
\s
\item{[12]} J.M. Murphy, Phys. Rev., {\bf D 8}, 4231 (1973).
\s
\item{[13]} D. Berkenstein and A. Meisels, Ap. J., {\bf 237}, 342 (1980).
\s
\item{[14]} E. Ruiz and J.M.M. Senovilla, Phys. Rev. {\bf
D45}, 1995 (1992).
\s
\item{[15]} A.K. Raychaudhuri, Private communication (1993).
\s
\item{[16]} A.K. Raychaudhuri, Phys. Rev. {\bf 90}, 1123 (1955).
\s
\item{[17]} C.G. Hewitt and J. Wainwright, Contribution in the Cape Town
workshop, June 27 - July 2, 1994.
\s
\item{[18]} R.C. Tolman,
Relativity, Thermodynamics and Cosmology (Oxford, 1934).
\s
\item{[19]} S. Weinberg, Gravitation and Cosmology (John Wiley, 1972).
\s
\item{[20]} N. Dadhich, L.K. Patel, K. S. Govinder and P. G.
L. Leach -- In preparation.
\s
\item{[21]} N. Dadhich and L.K. Patel, Preprint: IUCAA 20/93.
\s
\item{[22]} N. Dadhich and L.K. Patel, Preprint: IUCAA 21/93
(1993).
\s
\item{[23]} N. Dadhich, R. Tikekar and L.K. Patel, Current Science {\bf 65},
694 (1993).

\bye